\documentclass[a4paper,preprintnumbers,floatfix,superscriptaddress,prl,twocolumn,showpacs,notitlepage,longbibliography]{revtex4-1}
\usepackage[utf8]{inputenc}
\usepackage[T1]{fontenc}
\usepackage[sc,osf]{mathpazo}
\usepackage{amsmath, amsthm, amssymb,amsfonts,mathbbol,amstext}
\usepackage{graphicx}
\usepackage{dcolumn}
\usepackage{bm}
\usepackage{bbm}
\usepackage{hyperref}
\usepackage{mathtools}
\usepackage{comment}
\usepackage{color}
\usepackage{multirow}
\usepackage{diagbox}

\def\RR{\mathbbm{R}}

\def\1{\mathbf{1}}
\def\0{\mathbf{0}}

\def\st{\textrm{subject to }}

\def\p{\mathbf{p}}

\def\c{\mathbf{c}}

\def\t{\mathbf{t}}

\def\p{\mathbf{p}}

\def\q{\mathbf{q}}
\def\v{\mathbf{v}}

\newcommand{\ket}[1]{| #1 \rangle}
\newcommand{\bra}[1]{\langle #1 |}

\newcommand{\mean}[1]{\left\langle #1 \right\rangle}

\providecommand{\openone}{\mathbbm{1}}
\renewcommand{\rho}{\varrho}

\newcommand{\processnext}[1]{%
  \ifx\listfinish#1\empty\else\listact{#1}\expandafter\processnext\fi}

\DeclareGraphicsExtensions{.pdf,.png,.jpg}

\begin{document}

\title{Machine learning non-local correlations}

\author{Askery Canabarro}
\affiliation{International Institute of Physics, Federal University of Rio Grande do Norte, 59070-405 Natal, Brazil}
\affiliation{Grupo de F\'isica da Mat\'eria Condensada, N\'ucleo de Ci\^encias Exatas - NCEx, Campus Arapiraca, Universidade Federal de Alagoas, 57309-005 Arapiraca-AL, Brazil
}
\author{Samura\'i Brito}
\affiliation{International Institute of Physics, Federal University of Rio Grande do Norte, 59070-405 Natal, Brazil}
\author{Rafael Chaves}
\affiliation{International Institute of Physics, Federal University of Rio Grande do Norte, 59070-405 Natal, Brazil}
\affiliation{School of Science and Technology, Federal University of Rio Grande do Norte, 59078-970 Natal, Brazil}

\date{\today}
\begin{abstract}
The ability to witness non-local correlations lies at the core of foundational aspects of quantum mechanics and its application in the processing of information. Commonly, this is achieved via the violation of Bell inequalities. Unfortunately, however, their systematic derivation quickly becomes unfeasible as the scenario of interest grows in complexity. To cope with that, we propose here a machine learning approach for the detection and quantification of non-locality. It consists of an ensemble of multilayer perceptrons blended with genetic algorithms achieving a high performance in a number of relevant Bell scenarios. Our results offer a novel method and a proof-of-principle for the relevance of machine learning for understanding non-locality.
\end{abstract}

\maketitle
Quantum correlations, stronger than those allowed by classical systems, are at the core of quantum information science, its fundamental implications and practical applications \cite{Horodecki2009,Brunner2014}. For instance, the correlations obtained by measurements on distant entangled particles can violate Bell inequalities \cite{Bell1964}, not only proving the incompatibility of quantum theory with any local hidden variable (LHV) model but also paving the way to many relevant information processing tasks ranging from quantum cryptography \cite{Ekert1991,Barrett2005,Acin2007} and randomness certification \cite{Colbeck2007,Pironio2010} to self-testing \cite{Mayers2004} and distributed computing \cite{Buhrman2010}. To that aim, it is crucial to develop ways to test the incompatibility of a given correlation with LHV models, that is, to detect its non-local behavior.

The most common approach to that purpose is based on Bell inequalities. First, their violation is an unambiguous witness of the non-classicality of the correlations. Second, they serve as a objective function over which one can optimize quantum states and measurements to find violations and thus search for non-local correlations. Given its clear importance, over the years a very general framework has been developed \cite{Fine1982,Pitowsky1989} and dozens of inequalities were found \cite{Brunner2014}. LHV mo\-dels define a set of correlations compatible with it, the non-trivial boundaries of which are precisely the Bell inequalities. Typically, however, the characterization of the local set is computationally very demanding, rapidly becoming intractable as the scenario of interest raises its complexity \cite{Pitowsky1991}. Even the simplest Bell scenario, with two distant parties, cannot be fully characterized beyond a few particular cases where only a small number of measurements with few outcomes are allowed \cite{Collins2002,Collins2004,Brunner2008} . The situation is far worst for more general situations, for instance when dealing with quantum networks \cite{Branciard2010,Branciard2012,Chaves2016,Rosset2016,Tavakoli2016,Lee2018,Luo2018,Kela2017} where many independent sources of entangled states are present. In this case, LHV models give rise to semi-algebraic (non-convex) sets, the characterization of which has an even higher computational complexity \cite{Geiger1999,Garcia2005}. Faced with this impairing situation, it is natural to search for alternative routes to characterize and detect non-locality that do not rely on Bell inequalities. That is precisely the aim of this work.

Motivated by the outstanding recent progress within quantum physics problems \cite{Carrasquilla2017,PhysRevX.7.031038,1803.08823,Torlai2018,Gao2017,Carleo2017,Ma2017,Deng2018}, we propose here a machine learning (ML) approach to test non-classical behavior of correlations. Our starting point is to consider ``how far'' a given correlation is from the local set. As opposed to a specific Bell inequality -- covering a very limited region of the space of correlations-- our approach offers a global perspective of the local set geometry, in some sense testing all Bell inequalities at once.  We randomly sample the space of correlations and compute the distance to the local set by employing as a quantifier of non-locality the trace distance \cite{Brito2018}. This data is fed to an ensemble of deep learning algorithms \cite{Schmidhuber2015}, able to recognize patterns in the correlations and create models that not only can tell, with a high accuracy, whether a given point is local or not but also quantify its non-locality. Finally, by employing feature engineering and regression \cite{Zheng2018} we construct a ML objective function that can be optimized over to find new and relevant non-local points.

We show the relevance of our method by considering its application in a variety of Bell scenarios. In particular, the simplest scenario for which no complete characterization of the local set (Bell inequalities) is available. Further, we analyze an entanglement swapping experiment \cite{Zukowski1993,Carvacho2017,Saunders2017,Andreoli2017} giving rise to the notoriously thorny bilocality scenario \cite{Branciard2010,Branciard2012,Chaves2016}. Finally, we also show how the machine can learn to distinguish between quantum and post-quantum correlations \cite{Popescu1994,Navascues2007}, an important topic in the foundations of quantum theory \cite{Popescu2014}. 

\begin{figure}[t]
\begin{center}
\includegraphics[scale=0.3]{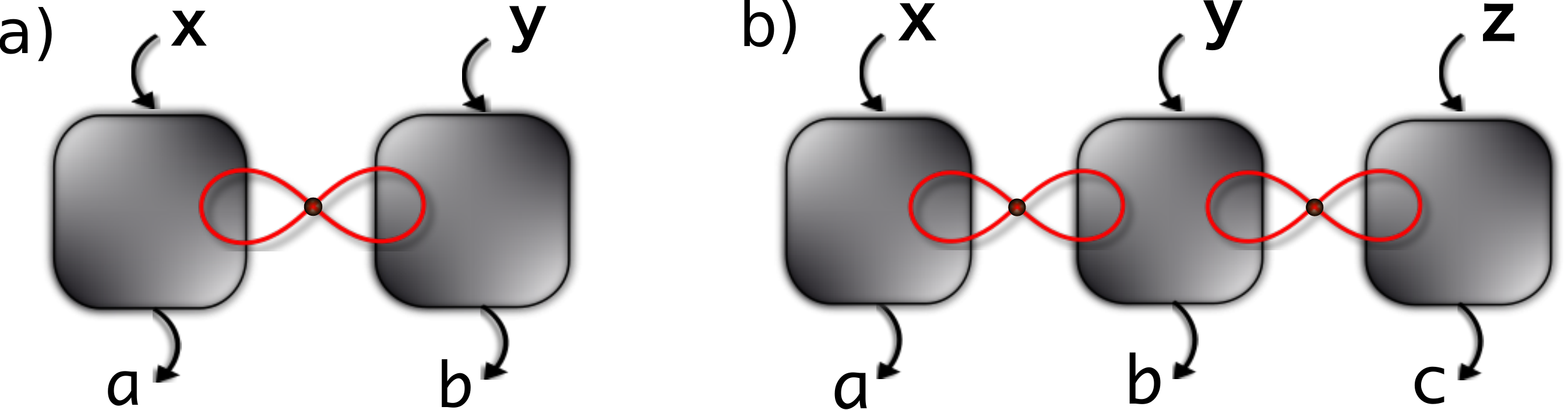}\vspace{0.8cm}
\includegraphics[scale=0.3]{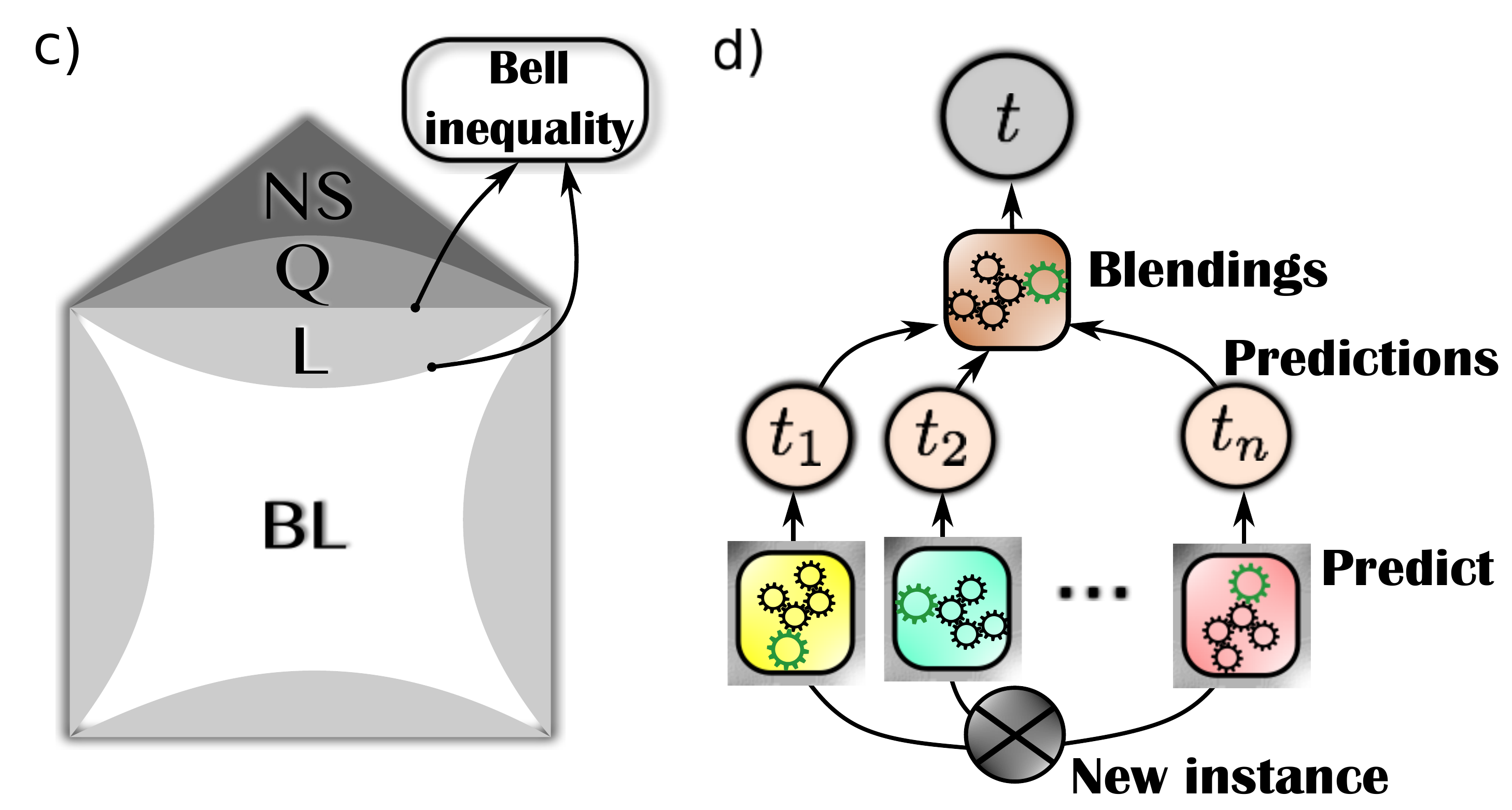}
\end{center}
\caption{Black-box representation of \textbf{a)} the bipartite Bell scenario and \textbf{b)} of a tripartite scenario with two independent sources of states. \textbf{c)} Pictorial illustration of the different sets of correlations: non-signaling, quantum, local and bilocal. \textbf{d)} Blending technique where different machines are combined to improve the overall performance.}
\label{fig_Bell}
\end{figure}

\emph{A machine learning approach to detect and quantify non-local correlations --}
Bell's theorem \cite{Bell1964} shows that measurements on distant entangled systems are incompatible with the assumption of local realism. We will refer to the simplest Bell scenario (see Fig.~\ref{fig_Bell}a), composed of two distant parties that, upon receiving their shares of a composite physical system, measure different obser\-vables (labeled by the variables $X$ and $Y$) obtaining a respective measurement outcome (labeled by $A$ and $B$). In a classical description, the probability distribution $p(A=a,B=b\vert X=x,Y=y)=p(a,b\vert x,y)$ observed in such a simple experiment should be decomposable in terms of a LHV model, that is,
\begin{equation}
\label{LHV}
p(a,b\vert x,y)= \sum_{\lambda} p(a\vert x,\lambda)p(b\vert y,\lambda) p(\lambda),
\end{equation}
defining a convex set $\mathcal{L}$, the boundaries of which are Bell inequalities (see Fig. \ref{fig_Bell}c). According to Born's rule, however, quantum mechanics implies that
\begin{equation}
\label{Quantum}
p(a,b\vert x,y)= Tr\left[ \left( M_{a}^{x}\otimes M_{b}^{y} \right)\rho_{AB} \right],
\end{equation}
where $\rho_{AB}$ is the density operator describing the shared physical system and $M_{a}^{x}$ and $M_{b}^{y}$ describe measurement operators. To test the non-locality of a given quantum distribution \eqref{Quantum} we thus have to show that it falls outside the set $\mathcal{L}$, the paradigmatic method for that being the violation of a Bell inequality.

However, the number of Bell inequalities grows very fast as the Bell scenario of interest grows its complexity (number of parties, measurements or outcomes) \cite{Pitowsky1989,Pitowsky1991,Brunner2014}, that is, any given inequality will typically offer very limited and localized information of a high-dimensional and intricate set of correlations. To cope with that, we employ here a more refined description, based on a non-locality quantifier $\mathrm{NL}(\q)$ given by minimum trace distance between the distribution $\q=q(a,b \vert x,y)$ under test and a $\p=p(a,b \vert x,y)$ in the set of local distributions \cite{Brito2018}:
\begin{eqnarray}
\label{NLtrace}
\mathrm{NL}(\q) =\frac{1}{2\vert x \vert \vert y \vert } \min_{\p \in \mathcal{L}} \sum_{a,b,x,y} \vert \q - \p \vert,
\end{eqnarray}
where $\vert x \vert =\vert y \vert=m$ denotes the number of possible measurement performed by the parties. 

Defined a Bell scenario of interest, the first step in our ML approach is to generate the training points to the machine. We do that by randomly sampling non-signalling (NS) distributions defined by simple linear constraints (see Appendix).
The reason for sampling NS instead of quantum distributions is three-fold. First, characterizing the quantum set is extremely challenging, the best available method given by a infinite hierarchy of semi-definite programs \cite{Navascues2007}. Second, even thought the NS condition allows for correlations beyond quantum mechanics, they play an important role in the foundations of the theory \cite{Popescu1994,Popescu2014}. Finally, as we will see, in spite of the machine being trained over the NS set, it provides a remarkable accurate description of the quantum set as well. To simplify the problem and without loss of generality, we do not use the full distribution as the input but rather the bipartite expectation value $\mean{A_xB_y}$.
For each sampled correlation we compute the corresponding distance measure and store this information as a vector $(\vec{f},t)$, where the components of $\vec{f}$ (known as features) stem for the different values of $\mean{A_xB_y}$ and $t=\mathrm{NL}(\q)$ (the target) \cite{FNote1}.

The samples are fed to different neural networks, the best ones been blended via a genetic algorithm \cite{Olson2016} to generate a final prediction for the target (see  Fig.~\ref{fig_Bell}d and Appendix). Following the standard approach, the data is split in a training and  cross-validation ($75 \%$) and test ($25 \%$) dataset, the first and second used to create a machine model generating a prediction $t_{\mathrm{pred}}$ and the second to test its accuracy in relation to the test targets $t_{\mathrm{test}}$. We also use a second test set generated by projective measurements on pure qubits states. To measure the performance/error of the model we employ the average trace distance $P= (1/N)\sum_{i=1}^N \vert \textbf{t}^i_{test}-\textbf{t}^i_{predicted} \vert$, where $N$ is the number of points in the testing set. 

\begin{figure}[t]
\begin{center}
\includegraphics[scale=0.35]{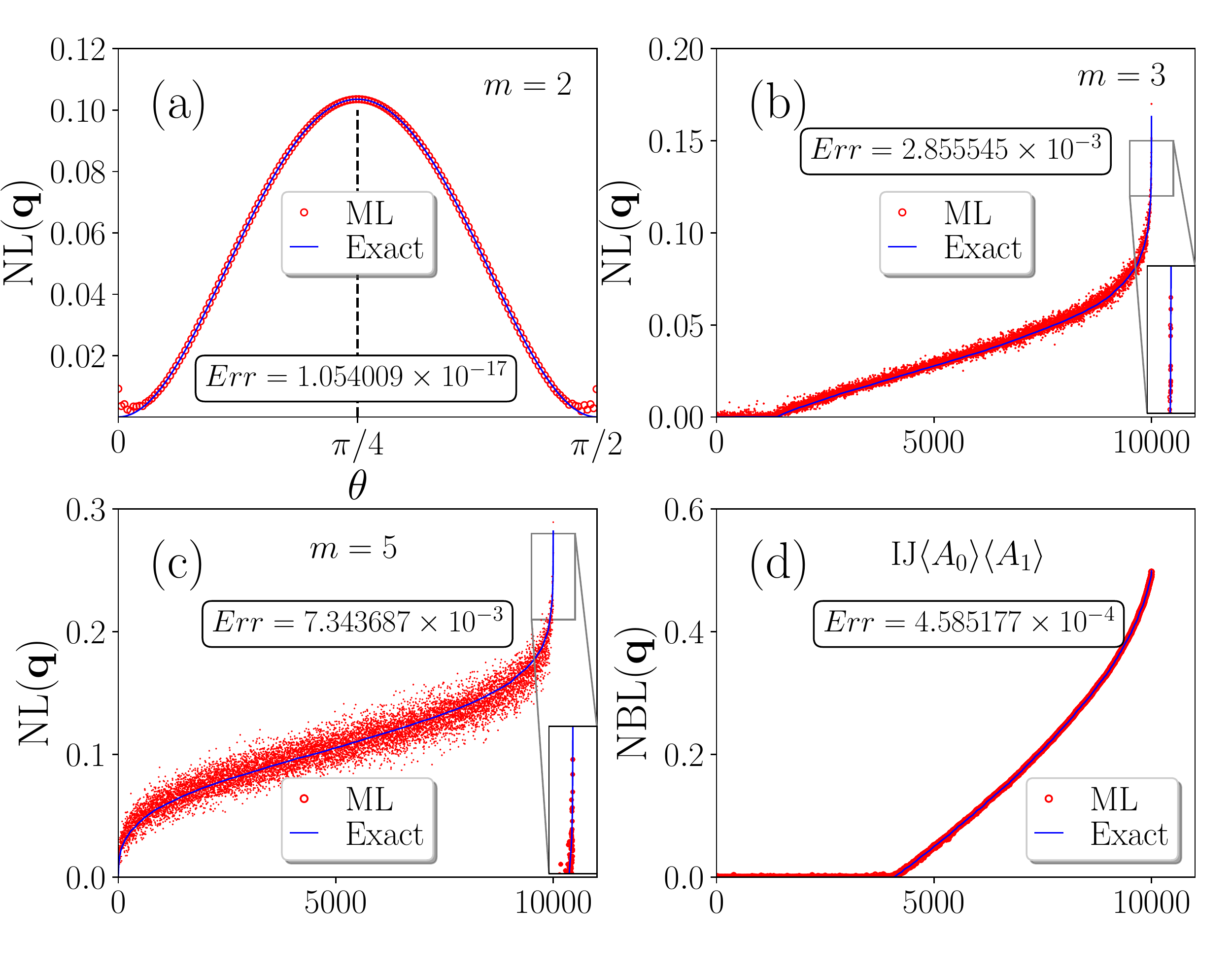}
\end{center}
\caption{In blue (straight line) the exact solution of Eq.~\ref{NLtrace} or Eq.~\ref{NBLTrace} and in red (circle) the ML prediction (considering $10^4$ test set points). In all cases the machine can predict, with excellent accuracy, the degree of non-locality without any information about Bell inequalities. \textbf{a)} Bipartite scenario ($m=2$) with quantum correlations obtained by projective measurements on $\ket\psi= \cos{\theta}\ket{00} + \sin{\theta}\ket{11}$, that maximally violate the $CHSH$ inequality \cite{Clauser1969}. \textbf{b)} Bipartite scenario ($m=3$), \textbf{c)} ($m=5$) and  \textbf{d)} the bilocality scenario employing 4 features $(\mathrm{I,J},\langle A_0\rangle, \langle A_1\rangle)$.}
\label{allresults}
\end{figure}

The results for the bipartite Bell scenario with $m=2,3,4,5$ dichotomic measurements and input data of $5 \times 10^5$ points are shown in Fig.~\ref{allresults} and Table~\ref{general_analysis}. The average error is of order $10^{-3}$ in all scenarios, both in NS and quantum test sets. The target function $\mathrm{NL}(\q)$ is a function of all Bell inequalities defining a given scenario and its number is equal to $8 (m=2)$, $72 (m=3)$, $27936 (m=4)$ while already for $m=5$ no complete characterization is available. Thus, such high accuracies are a truly remarkable feature of the deep learning approach. 

\emph{Non-local correlations in a simple quantum network--} Moving beyond the paradigmatic bipartite Bell scenario we consider the simplest possible quantum network, akin to an entanglement swapping experiment \cite{Zukowski1993}. It consists of three spatially separated parties interconnected by $2$ independent sources of quantum states (see Fig. \ref{fig_Bell}b). The LHV model describing such experiment implies that the observed distributions can be written as
\begin{eqnarray}
\label{bilocal}
& & p(a,b,c \vert x,y,z) = \\ \nonumber
& &\sum_{\lambda_1,\lambda_2} p(a\vert x, \lambda_1)p(b\vert y, \lambda_1,\lambda_2)p(c\vert z, \lambda_2)p(\lambda_1)p(\lambda_2).
\end{eqnarray}
\begin{table}[!h]
\caption{Performance (average norm-1) for different scenarios and different ML approaches. See Appendix for details.} 
\label{general_analysis}
\setlength\tabcolsep{0pt} 
\footnotesize\centering
\smallskip 
\begin{tabular*}{\columnwidth}{@{\extracolsep{\fill}}lcccccc}
\hline
\diagbox[width=8em]{Technique}{Scenario} & $m=2$ & $m=3$ & $m=4$ & $m=5$ & $\mathrm{IJA_0A_1}$ & $\mathrm{A_xB_yC_z,A_x}$\\ 
 \hline
 \hline
Typical MLP $(\times 10^{-3})$ & $0.46$ & $2.20$ & $7.75$ & $8.50$ & $2.70$ & $6.30$ \\
Blending $(\times 10^{-3})$ & $0.05$ & $1.54$ & $6.78$ & $7.31$ & $0.45$ & $3.22$ \\
\end{tabular*}
\end{table}

As opposed to the usual locality assumption, here we impose the independence of the two sources, that is, $p(\lambda_1,\lambda_2)=p(\lambda_1)p(\lambda_2)$, the bilocality assumption \cite{Branciard2010,Branciard2012}. Interestingly, there are local correlations that nonetheless are non-bilocal (see Fig. \ref{fig_Bell}c). In other terms, correlations that might appear of classical nature have their non-classicality revealed if the independence of the sources generating the correlations is taken into account. On the negative side, Eq. $\eqref{bilocal}$ defines an intricate non-convex set for which very few and specific inequalities have been derived so far \cite{Branciard2010,Branciard2012,Chaves2016,Rosset2016,Tavakoli2016,Lee2018,Luo2018}. To circumvent this difficulty we follow a similar approach to the one delineated before. Suppose, for instance that all the parties perform two possible measurements ($x,y,z=0,1$). Then the model \eqref{bilocal} implies the existence of a joint probability distribution such that the marginal between parts $A$ and $C$ factorize as $p(a_0,a_1,c_0,c_1)=p(a_0,a_1)p(c_0,c_1)$. A natural way to quantify the degree of non-bilocality is to measure how much a give correlation fails to comply with this statistical independence \cite{Chaves2015relax} and thus we consider as a measure the trace distance
\begin{equation}
\label{NBLTrace}
\mathrm{NBL}(\q)=\frac{1}{2}\min\sum_{a_0,a_1,c_0,c_1} \vert q_{a_0,a_1,c_0,c_1}-q_{a_0,a_1}q_{c_0,c_1} \vert,
\end{equation}
where the minimization is performed over all joint distributions $q(a_0,a_1,b_0,b_1,c_0,c_1)$ that marginalize to the distribution $q(a,b,c\vert x,y,z)$ under test. In spite of the non-convexity of the problem, $\mathrm{NBL}(\q)$ can be estimated via a sequence of linear programs (see \cite{Chaves2015relax} and Appendix for details).

\begin{figure}[t]
\begin{center}
\includegraphics[scale=0.25]{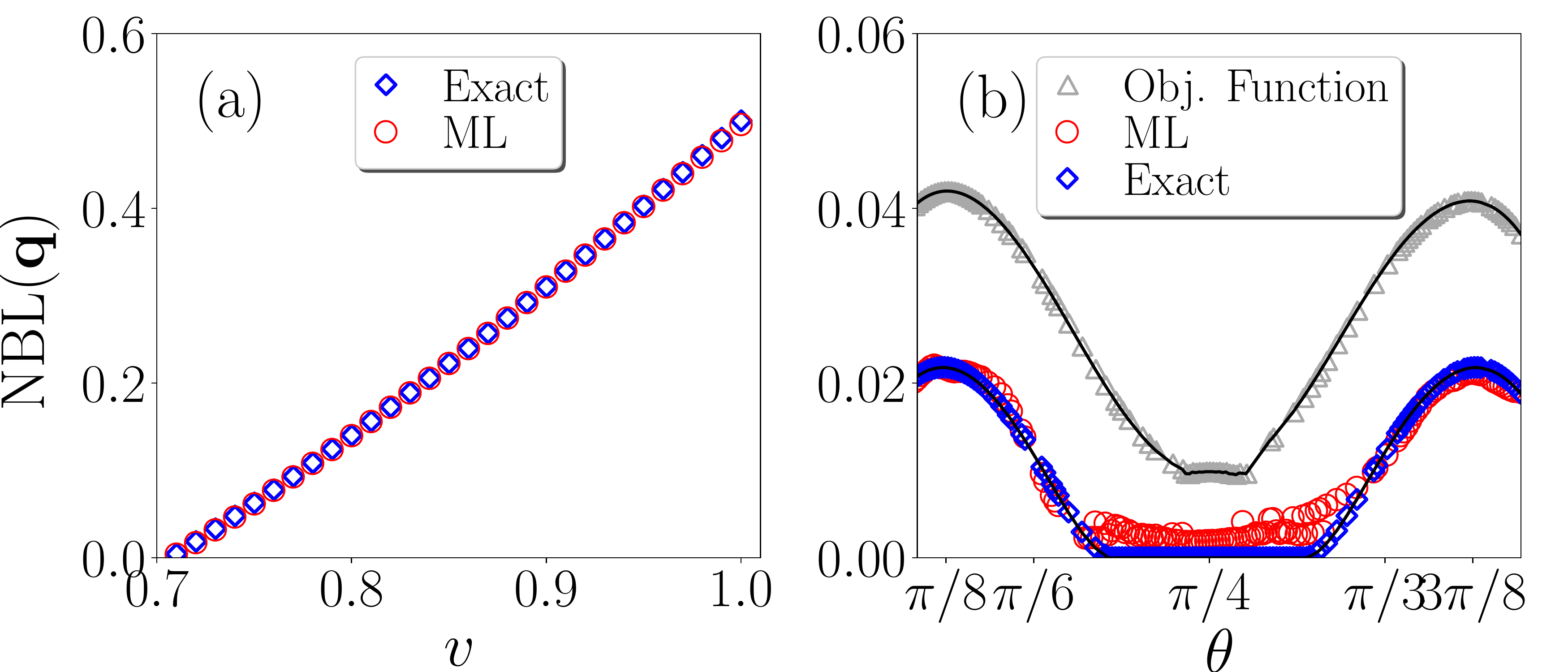}
\end{center}
\caption{\textbf{Bilocality scenario}. \textbf{a)} In diamond blue, the exact value of $\mathrm{NBL}= v^2 - 1/2$~\cite{Chaves2015relax} obtained by measurements on a Werner state maximally violating the inequality \ref{IJineq}. In red circle the deep learning prediction. \textbf{b)} In grey triangle, the results of a numerical optimization for the maximum value of the regression ML function for quantum correlations obtained by measurements on the state $\ket\psi_{AB}=\ket\psi_{BC} = \cos{\theta}\ket{00} + \sin{\theta}\ket{11}$ (not violating inequality~\ref{IJineq}).
In blue diamond the exact value and in red circle the prediction made by a neural network trained with NS correlations.
Strikingly, the ML approach can discover new quantum correlations without any information about Bell inequalities.}
\label{NBL_exact}
\end{figure}

We sample over NS distributions generating the input data $(\vec{f},t)$, with $t=\mathrm{NBL}(\q)$ and the features $\vec{f}$ encoding tripartite expectation values $\mean{A_xB_yC_z}$ and the marginal $\mean{A_x}$ \cite{Fnote2}. The known inequality in this scenario is given by
\begin{equation}
\label{IJineq}
\sqrt{\vert I \vert}+\sqrt{\vert J \vert } \leq 1,
\end{equation}
with $I=(1/4)\sum_{x,z} \mean{A_x,B_0,C_z}$ and $J=(1/4)\sum_{x,z} (-1)^{x+z}\mean{A_x,B_1,C_z}$.

We considered two scenarios, both considering a total of $1.5 \times 10^5$ points. In the first we considered $\vec{f}=(I,J,\mean{A_0},\mean{A_1})$ (4 features) and in the second $\vec{f}=(\mean{A_0B_0,C_0},\dots, \mean{A_1B_1,C_1},\mean{A_0},\mean{A_1})$ (10 features). The results are shown in Fig.~\ref{NBL_exact} and in Table \ref{general_analysis}. The overall performance is very high. Considering two situations, we have also compared the ML models trained with NS correlations to detect quantum ones. On the first, we have used the blended ensemble of deep learning models to compute the degree of non-bilocality of the correlations obtained by measurements on a Werner state $\rho=v \ket{\Phi^+}\bra{\Phi^+}+(1-v)\openone/4$ (with $\ket{\Phi^+}=(1/\sqrt{2})(\ket{00}+\ket{11})$) and that maximally violate the inequality \eqref{IJineq} (see Fig.~\ref{NBL_exact}a). On the second, we have numerically searched for quantum correlations violating the ML regression function but that do not violate \eqref{IJineq}. That is, in this case our ML approach is providing us with new and relevant information: the machine provides us new examples of correlations, the non-classicality of which cannot be detected by the known inequality \eqref{IJineq} (see Fig.~\ref{NBL_exact}b). We highlight that once the machine model is trained, to obtain a prediction about a new instance is basically instantaneous (of the order of $10^{-4}$ seconds) while the brute force method (used to train the machine) takes considerable more time, on average $20$ seconds; thus offering a $10^5$ speedup.

\emph{Machine learning post-quantum correlations --} The best available method to characterize the set of quantum correlations (those obtainable by measurements on a quantum state) is given by a hierarchy of semi-definite programs that converges asymptotically to the quantum set \cite{Navascues2007} and thus in general only provides an outer approximation. Notwithstanding, in some particular instances the convergence happens at at finite step, as it is the case in a bipartite scenario where each party can perform two possible dichotomic measurements.

\begin{table}[!h]
\caption{The confusion matrix $C_{ij}$ of the blend of classifiers for $10^5$ unseen inputs, which returns the number of observations known to be in group $i$ but predicted to be in group $j$. The sum of the elements of the main diagonal divided by the total of elements gives the accuracy score.} 
\label{confusion_matrix}
\setlength\tabcolsep{0pt} 
\footnotesize\centering
\smallskip 
\begin{tabular*}{\columnwidth}{@{\extracolsep{\fill}}lccccc}
\hline
\diagbox[width=10em]{True Class}{Predictions} & Local & Quantum  &  Post-quantum\\  
  \hline
  \hline
 Local & $33436$ & $96$ & $0$ \\
 Quantum & $41$ & $33173$ & $236$ \\
 Post-quantum & $0$ & $136$ & $32882$ \\
  \end{tabular*}
\end{table}

A necessary and sufficient condition \cite{Masanes2003} for the expectation values $\mean{A_xB_y}$ with $x,y=0,1$ to have a quantum realization is given by all four symmetries of the inequality $\vert \arcsin{\mean{A_0B_0}}+\arcsin{\mean{A_0B_1}}+\arcsin{\mean{A_1B_0}}-\arcsin{\mean{A_1B_1}} \vert \leq \pi$. Furthermore, the non-locality of the associated distribution can be also be decided by testing all the symmetries of the inequality $\vert \mean{A_0B_0}+\mean{A_0B_1}+\mean{A_1B_0}-\mean{A_1B_1} \vert \leq 2$.  Given the list of correlators $\mean{A_xB_y}$ we can then classify it as local, non-local (quantum) or post-quantum. 

In machine learning, classification is the problem of determining to which class of categories a new observation belongs, by means of a training set of data containing instances whose category membership is known \cite{aaron}. The ensemble of classifiers, created in a similar way as the ensemble of regressors, was trained over $4 \times 10^5$ inputs points. The overall accuracy achieved was $99.49 \%$. To better quantify the quality of the predictions of the ensemble of deep learning models that we proposed, we computed the confusion matrix for a random sample of $10^5$ unseen new instances in Table \ref{confusion_matrix}, see Appendix for more details. Interestingly, even though there are many post-quantum points close to the local set, the ML method never make mistakes between both. 

\emph{Discussion-- } Bell non-locality shows that even without a precise description of a physical apparatus and solely based on measurement data, one can prove the quantumness of some observed correlations. It is at the core of the device-independent approach to quantum information processing \cite{Pironio2016} with many applications in near term quantum technologies such as quantum criptography \cite{Ekert1991,Barrett2005,Acin2007}. Detecting Bell non-locality beyond simple cases, however, remains an thorny issue given the hard computational complexity of the characterization of locality via Bell inequalities \cite{Pitowsky1989,Pitowsky1991}. Further, with the recent advances on the quantum internet \cite{Kimble2008,Castelvecchi2018} --in short, a network with several independent sources of quantum states-- such computational difficulties become even more pronounced \cite{Geiger1999,Garcia2005,Chaves2016}. Here we propose an alternative and timely route, a machine learning approach, allowing the detection and quantification of non-locality as well as its quantum (or post-quantum) nature. To illustrate its benefits we have applied it to a number of relevant Bell scenarios showing that not only the machine can learn but also teach, for instance pointing to new kinds of non-local correlations that cannot be detected by known Bell inequalities.

Our results provide a proof-of-principle for the relevance of ML tools in Bell non-locality and we trust will open several research venues. A natural next step is to consider classical and quantum networks of growing complexity \cite{Chaves2016,Rosset2016,Tavakoli2016,Lee2018,Luo2018,Kela2017}. Another clear possibility is the combination with other recent results, e.g., the reinforcement learning approach to find the maximum violation of a given Bell inequality \cite{Deng2018}.

\section*{Acknowledgements}
The authors acknowledge the Brazilian ministries MEC and MCTIC, funding agency CNPq (AC's Universal grant No. 423713/2016-7, RC's PQ grant No. 307172/2017-1 and INCT-IQ) and UFAL (AC's paid license for scientific cooperation at UFRN).

\bibliography{mybibfile}

\begin{thebibliography}{59}
\bibitem [{FNo()}]{FNote1}%
  \BibitemOpen
  \href@noop {} {\ }\bibinfo {note} {The measure $\mathrm{NL}(\q)$ is optimized
  over all $q(a,b\vert x,y)$ compatible with a given
  $\mean{A_xB_y}$.}\BibitemShut {Stop}%
\bibitem [{Fno()}]{Fnote2}%
  \BibitemOpen
  \href@noop {} {\ }\bibinfo {note} {The marginal information provided by
  $\mean{A_0}$ is essential to compute $\mathrm{NBL}(\q)$. Please see Appendix
  for more details.}\BibitemShut {Stop}%
\end{thebibliography}

\section{Appendix}

\subsection{Machine and Deep Learning Overview}

Machine learning can be defined as automated processes that retrieve patterns and/or relations from data without being explicitly programmed to. By means of statistical techniques, computers can improve their performance $p$ in solving a task $T$ by being exposed to examples or experiences $E$. That is, machine learning happens whenever $p(T) \propto E$. Here, we provide a succinct but insightful description of all machine learning steps involved in our work. These include the tasks (regression and classification), the experiences (supervised learning), the machine learning approaches and algorithms (such as the multilayer perceptron) and also the performance measures one can use (trace norms, confusion matrices, etc). For more details see \cite{aaron}.

Overall, machine learning is a method used to construct complex models to make predictions in problems hard to solve with fixed programs. In principle, it may as well shed new light on how intelligence works. However, this is not, in general, the main purpose behind machine learning techniques given that by "learning" it is usually meant the skill to perform the task better and better, not necessarily how it is learning the task itself. For instance, for companies like Google it is enough to know that its huge neural network is classifying very well the binary "spam/not spam" problem, not paying much attention in how exactly it is doing it. Often the machine learning model is seen as a “black-box”, hard or even impossible to interpret, simply satisfying ourselves with the answer provided by the machine. 

\subsubsection{Tasks: classification and regression}

Quite generally, a machine learning task is specified by how the machine process a given set of input data sampled from the problem/system at hand. Each instance of data consists of a few features and can represented by a vector $\mathbf{X}_i$. All instances of input data $\mathbf{X}_i$ (vectors by itself) are encoded in another vector $\mathbf{X} \in \mathbb{R}^{n}$. Amid the most common machine learning tasks we can cite: classification, regression, transcription, machine translation, anomaly detection, among many others. Below we briefly describe the two kinds of tasks we employ in this work: classification and regression.

\subsubsection{1. Classification}

In this task the program must specify in which of $k$ possible categories a given input instance should be classified. In this manner the algorithm is requested to develop a function $f: \mathbb{R}^{n} \to \{1, ..., k\}$, where $k$ is a finite and (typically) pre-established integer number. Therefore, given an input vector $\mathbf{X}$, the model returns a numeric value (the target) $t = f(\mathbf{X})$. In general, $f$ returns a normalized probability distribution over the $k$ classes and the suggested class is the one with highest probability. Referring to the main text, this is the case, for instance, when we desire to classify the given Bell correlations as local, quantum or post-quantum.       

\subsubsection{2. Regression}

Here, the computer program is requested to predict a numerical real value (the targer) $t$ to some input. Therefore, the algorithm models a function $f: \mathbb{R}^{n} \to \mathbb{R}$. This can be considered a similar task as the classification with just a distinct output format as, for example, in the cases that we predict the distances to the bi-local  and local sets.

\subsubsection{The experience E: supervised learning}

Learning algorithms are commonly classified as unsupervised and supervised, depending on the way the learning process occurs given a collection of data points (dataset). The techniques used in this work are called supervised learning, once the learner experiences a dataset of features $\mathbf{X}$ and also the target or label vector $\mathbf{y}$, provided by a "teacher", hence the term "supervised".

In other words, the learner is presented with example inputs and their known outputs and the aim is to create a general rule that maps inputs to outputs, by generally estimating the conditional probability $p(\mathbf{y} | \mathbf{X})$. In unsupervised learning, as there is no teacher, the learner must fathom by himself how to deal with the data. 

\subsubsection{The performance p}

One central aspects of machine learning which differentiates it from an optimization approach is that we want the learner to perform well on unseen inputs, what is called generalization. To accomplish this, we compute the performance in a set previously separated from the data we use to train the machine, named test set, which usually corresponds to $20 \%-25 \%$ of the available data.  

For classification tasks, one useful measure is the accuracy score which corresponds to the rate of correct predictions produced by the model. One similar performance measurement is the error rate, which retrieves the proportion of the incorrect predictions. For instance, for the classification task in this work, we showed the ensemble of machines showed a accuracy of about $99.49 \%$, meaning an error rate of $0.51 \%$.  For a multi-classification task as the one we investigated, the confusion matrix is a more insightful measure, once it allows to see where the machine faces more difficulties (see the corresponding section for more details). 

For regression tasks, a measure of accuracy as above stated is no longer viable. Often, one evaluates the mean absolute error (MAE) or the mean square error (MSE), which corresponds to the mean L$1$ and L$2$ norms, respectively. Although they can be indistinguishably used for the evaluation of the generalization error on the test set with no great influence on the model, it plays a crucial role when used on the training set due to the optimization search performed there. For instance, if we try to minimize a cost function which looks like a MAE, a sequence of $n$ errors of the order $\epsilon$ is equivalent to a single large error of size $n \epsilon$, meaning that in your project a large number of medium-size errors is as much as acceptable as a few larger errors. However, if a L$2$-like cost function is used the model is calibrated to accept medium-size errors throughout the learning process, but not large errors. 

Therefore, the ideal cost function varies from project to project. For our purpose of predicting the distance to the local or bi-local sets, it is optimal to prevent large errors once we aim to unveil the  target value with sufficient precision regardless of its actual value. In this manner, we used a L$2$-like cost function which is in fact the default cost function of the Python scikit-learn package for the implementation of the multilayer perceptrons we used \cite{scikit}. However, to present the error evaluation on the test set, we choose to present the L$1$ error due to its straightforward interpretation as quantifying the degree of non-locality (see eqs. \eqref{NLtrace} and \eqref{NBLTrace}).   

\subsubsection{Multilayer Perceptron (MLP)} 

Multilayer Perceptrons are the backbone of deep learning modeling, belonging to the class of artificial feed-forward neural networks with the main goal of approximating a function $f(\mathbf{X})$ by $f^{*}(\mathbf{X};\overrightarrow{\theta})$ which maps an input $\mathbf{X}$ to an output $\mathbf{y}$ returning the best values of the parameters $\overrightarrow{\theta}$ after the learning process.

\begin{figure}[ht!] \centering
\includegraphics[scale =0.35]{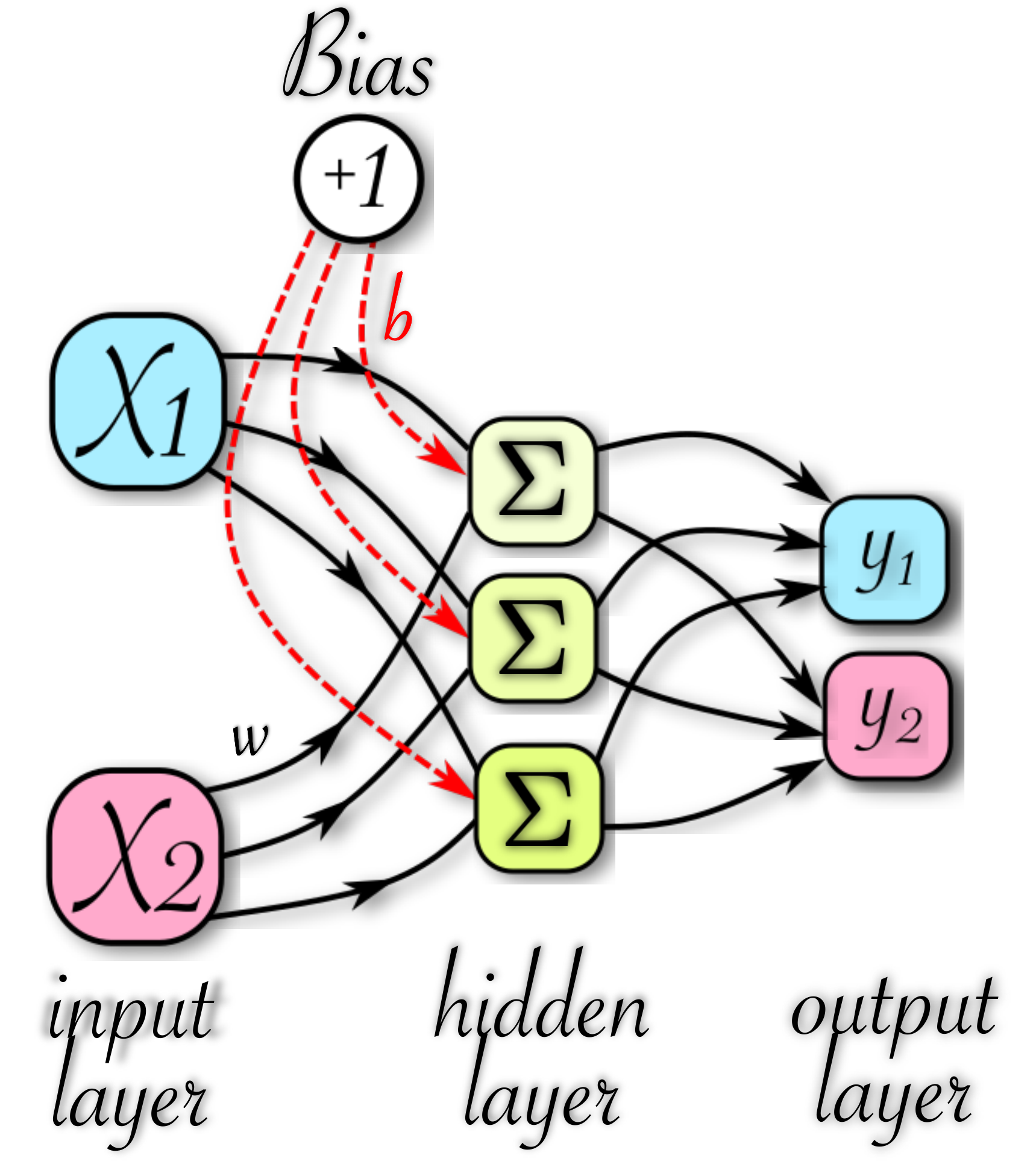}
\caption{Schematic representation of an artificial neural network composed of three layers with only one hidden layer. The nodes represent neurons and the solid (dashed) arrows stand for weight (bias) among neurons. Each neuron processes the incoming signals by means of a activation function $\Sigma$.}
\label{f27}
\end{figure}

Originally conceived to artificially reproduce the functionality of a central nervous system composed a tantamount of highly connected neurons layers (hence the term "neural") in the task of pattern recognition, MLPs are still one of the most powerful machine learning approaches for complex tasks. They are said to belong to the feed-forward class due to the unidirectional flux of information from $\mathbf{X}$ to $\mathbf{y}$, or in other words, from the input layer to the output layer as schematically shown in Fig. \ref{f27}. Each layer $i$ can be considered a function $f^{(i)}$ and the model is therefore a network or composition of functions, $i$ representing the depth of the deep learning machine. They are also fully connected as every neuron of a given layer is connected to every neuron of the next layer.

Roughly speaking, a given neuron receives various signals from the other neurons and "decides" if it should activate or not by means of an activation function $\Sigma$ which adds up all the incoming contributions. In early works inspired by neuroscience, a common choice was the standard logistic function $\Sigma (x) = 1/(1 + e^{-x})$, however in modern approaches the choice follows the direction that optimizes the predictions both in precision as well as in computational time. Throughout this work, we used the rectified linear unit (ReLU) function, $\Sigma(x) = \text{max}(0, x)$. The middle layers are said to be hidden layers as it is not known what they must pass for the next layer in order to achieve the general purpose of outputting a value $y$ close to $f(X)$ for each $X$. 

Once a cost function $J(\overrightarrow{\theta})$ is defined, for instance the MSE
\begin{eqnarray}
J_{\text{MSE}}(\overrightarrow{\theta}) = \frac{1}{m} \sum_{\mathbf{x} \in \mathbf{X}^{\text{train}} } (\mathbf{y}^{\text{train}} - f^{*}(\mathbf{x};\overrightarrow{\theta}) )^2,
\end{eqnarray}
where $m$ is the training set size, one might face the very difficult problem of computing the gradient, $\nabla_{\overrightarrow{\theta}} J(\overrightarrow{\theta})$, of a highly complex function, because even a model with a few hidden layers of hundreds of neurons can have thousands or millions of parameters. The optimal $\overrightarrow{\theta}$ returns the values of the weights $\bf{w}$ (solid arrows in Fig. \ref{f27}) and the bias $\bf{b}$ (dashed arrows in in Fig. \ref{f27}) which generalizes better. The bias acts somewhat as the non-null intercept in a linear regression problem, amplifying the possible solutions. 

In recent deep learning approaches, one overcomes this computational cost by minimizing the error in the direction from the output layer to the input layer with the back-propagation algorithm and its variants. In our codes, we used the \textit{Adam} algorithm natively contained in the scikit-learn package for MLP, see Ref. \cite{adam} for more details.

\subsection{Ensemble Learning}
\label{sec:ensem}

When you combine the predictions of a committee of predictors (classifiers or regressors), you will likely get better predictions than with the best individual predictor. A group of predictors is called an ensemble. 

In ensemble learning methods one strategically combines various machine learning algorithms aiming to improve the prediction performance. This improvement is feasible due to some advantages the ensemble delivers. One clear advantage is statistical. The ensemble reaches a higher precision with less training data as one can see, for example, in Fig.~\ref{MAE_232}. It was needed only $60 \%$ of the data (300k points) to achieve the same precision of a single MLP with a 500k dataset. The second justification is computational. As neural nets, for instance, can retrieve locally optimal answers due to local minimum lockups, an ensemble of MLPs can be used to discard aberrant solutions. Another reason is that the true function $f$ may not be well represented with just one model or hypothesis, therefore the aggregation of many hypotheses can deliver a better approximation \cite{ensemble}.  

Most ensemble methods use algorithms of the same type leading to homogeneous ensembles as the one we propose, an ensemble of just MLPs. For ensemble methods to be better than any of its individual members, the learners must be as precise and as diverse as possible. In section \ref{sec:struc}, we present details of how we dealt with the precision and diversification of our ensemble. 

The ensemble method we use here is called stacking, where a new model is trained to aggregate the predictions of all predictors instead of using simple functions, such as averages. The final predictors is named the blender. See Fig. \ref{fig_Bell}d for a schematic illustration of the ensemble performing a prediction task on a new instance. Our novel contribution was to use a genetic algorithm to search for the ideal blender.

\subsubsection{\label{sec:tpot}Genetic programming: the TPOT tool}

As there is no \textit{a priori} machine learning algorithm more suitable for the task at hand, according to the well-known No Free Lunch Theorem, when implementing a ML project one has to test a myriad of ML approaches with a vast hyper-parameter space to cover in order to find a suitable model.  

TPOT is a free Automated Machine Learning (AutoML) tool written in Python that can optimize machine learning pipelines using genetic programming. It automates the exhaustive stage of exploring various possible pipelines to find the optimal one for the data provided \cite{tpot}. 

In a nutshell, for a regression or classification task and providing only the raw input and output data, it automatically tests many common stages in ML application, e. g., pre-processing (checks data type and applies normalization), feature engineering (looks if higher powers of the input data is relevant), dimension reduction (performs principal component analysis (PCA) if needed), hyperparameter optimization and so on, returning an optimal pipeline. However, so far, it does not contain deep learning algorithms in its search for the best solution. When we tried this approach for complex scenarios like $(2,4,2)$ and $(2,5,2)$, the best pipeline returned a similar MAE as of a high-order polynomial fit, therefore not giving a adequate solution. Withal, it indicates that in our optimal pipeline we should include a polynomial feature engineering of degree $2$ and that no normalization or dimension reduction (such as Principal Component Analysis) were needed,  giving a considerable advantage in implementing deep learning approaches.   

\subsubsection{\label{sec:struc}The Ensemble Structure}

We initiated by verifying that no dimension reduction was needed according to the output of the TPOT tool. Taking, for instance, the bilocality problem with $10$ features, we used scikit-learn PCA analysis to investigated the features importance and we got the following vector [$0.10610683$ $0.10405454$ $0.1016126$  $0.10085203$ $0.09957134$ $0.09925612$
 $0.09897048$ $0.09820423$ $0.09754938$ $0.09382245$], meaning that all the features have similar relevance and the data cannot be projected into a lower dimensional space without losing much information, confirming the TPOT suggestion. This is totally compatible with what one should expect, as these features correspond to $10$ independent expectation values.
 
Also, this motivated us to search for the deep learning approach which has less implementation cost, being precisely the MLP. As the input features display no correlations among them, more powerful (and costly) methods such as Convolutional Neural Networks would not, in principle, give a considerable larger precision. 

\subsubsection{Training the MLPs}

The only pre-processing we have done was a polynomial feature engineering of degree two before delivering the input data to the training stage, as suggested by the TPOT tool. It consist in generating a new feature matrix consisting of all polynomial combinations of the features with degree less than or equal to the specified degree. For instance, for a two features input $[a, b]$, the degree-2 polynomial features are $[a, b, a^2, ab, b^2]$, bias excluded as it can be retrieved by the MLP.

We propose an ensemble method composed of independent MLPs trained with distinct number of layers and neurons blended with a TPOT solution. The overall procedure can be summarized as: 

$1)$ We independently train $36$ MLPs with the number of layers ranging from $2$ to $5$ and the quantity of neurons ranging from $100$ to $500$, augmenting by $50$ in each step. This covers the diversification demand for the ensemble to work properly, as discussed in related section.

We choose to start with at least $2$ hidden layers once it represents the composition of $2$ functions, \textit{a priori} dealing better with high nonlinearities. If the task is a classification (regression) we used the corresponding scikit-learn module MLPClassifier (MLPRegressor), respectively. For all of them we use the same parametric configuration: gradient solver (Adam), learning rate ($10^{-5}$), activation function (ReLU), as it is an empirical successful choice for many applications. The training is done by minimizing a MSE cost function due to the fact that it allows many small-medium size errors, but forbids individual larger ones, as already discussed. They are, in fact, default parameters in the package. 

$2)$ After training and validating each MLP by checking that the generalization error is of the same order as the training error, we exclude the machines whose test set MAE is great or equal than $70 \%$ of the MAE of a polynomial fit of degree-$4$ over the data and collect the remaining "opinions" of the best machines. For the classification task, the base accuracy was set to be $98.5 \%$, as proposed by the direct TPOT optimal solution. In this way, all the classifiers with accuracy below this threshold is rejected. These procedures renders the "good" precision requirement we discuss in ensemble learning section. 

$3)$ We propose that instead of just taking the average of the predictions $t_i$ of each MLP, we pass this to a TPOT layer, responsible to perform the blend to achieve the ultimate best prediction $t$ for a given input $X$, see Fig. \ref{allresults} (d). 

This stage of requesting the genetic algorithm to find the best approach to the final answer also serves to generalize our framework to deal with regression as well as classification tasks. For instance, for the multi-classification task we dealt, simply taking the average of the opinions of four machines where two output 1 (local) and the other two output 3 (post-quantum) would result, on average, in a final answer stating 2 (quantum) as a solution, which is very unlikely to be the best guess.

For the regression tasks, we choose to search for the TPOT solution for the most complex bipartite Bell scenario ($m=5$) composed of $25$ features and applied the suggested solution to all the others scenarios of both bilocality and nonlocality investigation. The TPOT optimal regressor model was a Gradient Boosting Regressor. For the classification task, the suggested solution was an Extra Trees Classifier. For further details about these techniques, the reader is suggested the Refs. \cite{scikit,ert}. 
In Table \ref{general_analysis}, it is shown how the ensemble typically outperforms an individual MLP (in some cases achieving order of magnitude improvements), being a better approach specially for more complex problems. In the next subsection we discuss in detail the confusion matrix to highlight how the ensemble of classifiers outperform a single MLP classifier.  

\subsubsection{Confusion Matrices}

A confusion matrix, also known as an error matrix, is a specific table layout that allows visualization of the performance of an algorithm, typically a supervised learning one. Each row of the matrix represents the instances in a predicted class while each column represents the instances in an actual class. The name stems from the fact that it makes it easy to see if the system is confusing two classes, commonly mislabeling one as another \cite{aaron}.

Here, we present the confusion matrix and accuracy scores of a single MLP so that we can compare with the result produced by the blending of a ensemble of MLP in Table \ref{confusion_matrix}.

\begin{table}[!h]
\caption{The elements of the confusion matrix $C_{ij}$ of a single MLP for the same $10^5$ as in Table \ref{confusion_matrix}, which return the number of observations known to be in group $i$ but predicted to be in group $j$. The sum of the elements of the main diagonal divided by the sum of total of elements gives the accuracy score.} 
\label{confusion_matrix_single}
\setlength\tabcolsep{0pt} 
\footnotesize\centering
\smallskip 
\begin{tabular*}{\columnwidth}{@{\extracolsep{\fill}}lccccc}
\hline
\diagbox[width=10em]{True Class}{Predictions} & Local & Quantum  &  Post-quantum\\  
  \hline
  \hline
 Local & $32885$ & $238$ & $0$ \\
 Quantum & $60$ & $32951$ & $469$ \\
 Post-quantum & $0$ & $406$ & $32991$ \\
  \end{tabular*}
\end{table}

The accuracy score of a typical single MLP is about $98.83 \%$, check Table \ref{confusion_matrix_single} for details. The blending accuracy is $99.49 \%$, therefore reducing the error rate from $1.17 \%$ to $0.51\%$, yielding an improvement of more than $56 \%$. Furthermore, comparing Tables \ref{confusion_matrix} and \ref{confusion_matrix_single}, we notice a similar reduction for the misclassification between quantum/post-quantum and local/quantum cases, meaning that the ensemble is improving the performance equitably.

It worth mentioning that we provide the same amount of local, quantum and post-quantum examples. Therefore, in each split, there is roughly the same number of 
samples from each class, meaning a balanced classification problem, which yields a non-biased classification accuracy.

\subsubsection{Improving the precision with larger training sets}

\begin{figure}[ht!]
\begin{center}
\includegraphics[scale=0.35]{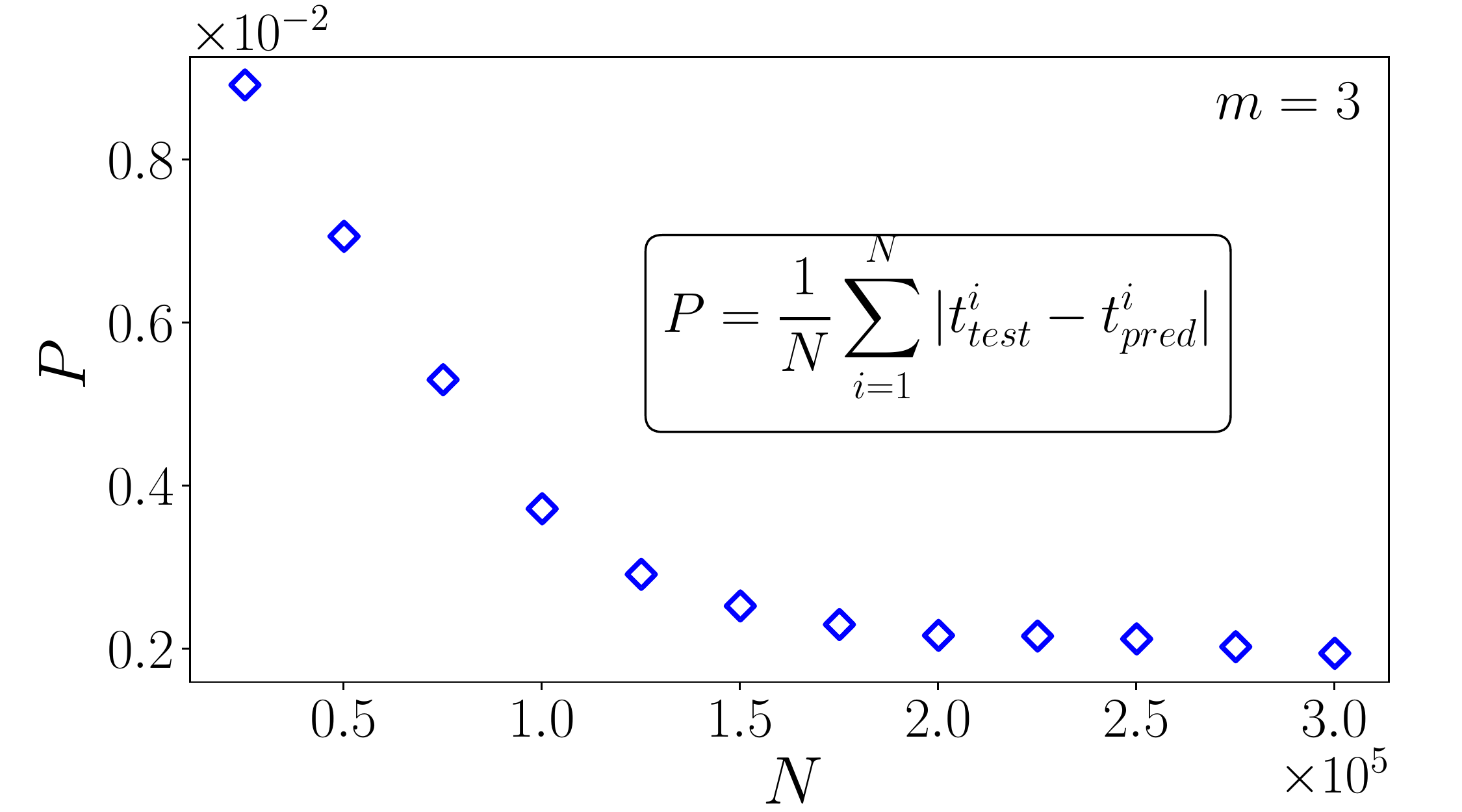}
\end{center}
\caption{Average trace distance error for $m=3$ bipartite Bell scenario versus size $N$ of the training set.}
\label{MAE_232}
\end{figure}

Typically, the more training points are provided the better will the precision of the ML model over the unseen instances. However, for a fixed neural networks (number of layers and neurons), typically there will also exist a plateau beyond which no improvements will be made by increasing the number of training points. This is clearly shown in Fig.~\ref{MAE_232} considering the bipartite Bell scenario with $m=3$. Training sets with more than $1.5 \times 10^5$ do not improve significantly the precision. This is the stage when one needs more powerful models, such as the blend approach we propose.

\subsection{Linear program formulation}
\subsubsection{\bf{A.} $\mathrm{NL(\q)}$ computation}
The $\mathrm{NL}(\q)$ quantifier gives us the minimal distance of $\q$ from the local set of correlations. Given a distribution $\q=q(a,b \vert x,y)$ of interest, in order to compute $\mathrm{NL}(\q)$ we have to solve the following optimization problem
\begin{eqnarray}
\min_{\lambda \in \RR^m} & & \quad \| \q - A \cdot \lambda \|_{\ell_1} \\ \nonumber
\st & & \quad \lambda \geq 0 \\ \nonumber 
 & & \quad \sum_{i}\lambda_i=1,
\end{eqnarray}
where the classical correlations are defined by $\p_C=\mathrm{A}\cdot \vec{\lambda}$, with $\lambda$ being a vector with components $\lambda_i = p(\lambda = i)$ and $A$ being the matrix with entries $A_{j,i} = \delta_{a,f_a(x,\lambda=i)} \delta_{b,f_b(y,\lambda=i)}$, where $j=(a,b,x,y)$ and $f_a$ and $f_b$ are deterministic functions. Once that $\ell_1$-norm optimization problem can be written as
\begin{eqnarray}
\|  \q \|_{\ell_1} = \min_{\t \in \RR^n} & & \quad \langle \1_n, \t \rangle   \label{L1_LP}\\ \nonumber
\st & & \quad -\t \leq \q \leq \t \\ \nonumber
& & \quad \q\geq \0_n,
\end{eqnarray}
adding the appropriated constraints for our problem, the linear problem becomes
\begin{eqnarray}
\label{NLLP2}
\min_{\t \in \RR^n, \vec{\lambda} \in \RR^m, \q \in \RR^n} & & \quad \langle \1_n, \t \rangle   \\ \nonumber
\st & & \quad -\t \leq \q - A \cdot \vec{\lambda} \leq \t \\ \nonumber
 & & \quad \sum_{i}\lambda_i=1 \\ \nonumber
 & & \quad \sum_{a,b}\q(ab|xy) = 1 \\ \nonumber
 & & \quad \sum_{a}\q(ab|xy) -\sum_{a}\q(ab|x'y)= 0 \;\;\forall\, (b,y)\\ \nonumber
 & & \quad \sum_{b}\q(ab|xy) -\sum_{b}\q(ab|xy')= 0 \;\;\forall \,(a,x)\\ \nonumber
   & & \quad \mathrm{M}_{cor}\cdot \q = \v_{cor} \\ \nonumber
 & & \quad \vec{\lambda} \geq \0_m \\ \nonumber
 & &\quad \q \geq \0_n. \\ \nonumber  
\end{eqnarray}
Where $\q$ is a vector of probability distribution to which we want to quantify the non-locality and $\v_{cor}$ is the set of full correlators given by $\v_{cor}=[\langle A_0B_0\rangle,...,\langle A_{m-1}B_{m-1}\rangle]$. Since we are working with correlators (expectation values), we need to put the constraint $\mathrm{M}_{cor}\q = \v_{cor}$ in our LP, where $\mathrm{M}_{cor}$ is the transformation matrix constructed from the relation $\mean{A_xB_y} = \sum_{a,b=0}^{1}(-1)^{a+b} q(a,b \vert x,y)$ for $a,b = 0,1$ and $x,y =0,1,...,m-1$. The vector $\v_{cor}$ has dimension $m^2$ and each component is a value  randomly chosen in the interval $[-1,1]$.

\subsubsection{\bf{B.} $\mathrm{NBL(\q)}$ computation}
The $\mathrm{NBL}(\q)$ quantifier gives us the minimal distance of $\q$ from the bilocal set of correlations. Bilocal correlations are a non-convex set that obey
\begin{eqnarray}
\label{bilocal}
& & p(a,b,c \vert x,y,z) = \\ \nonumber
& &\sum_{\lambda_1,\lambda_2} p(a\vert x, \lambda_1)p(b\vert y, \lambda_1,\lambda_2)p(c\vert z, \lambda_2)p(\lambda_1)p(\lambda_2),
\end{eqnarray}
where we assume that the two sources are independent $p(\lambda_1,\lambda_2) = p(\lambda_1)p(\lambda_2)$. To compute the $\mathrm{NBL(\q)}$, the following minimization problem needs to be solved:
\begin{eqnarray}
\label{NBLLP}
\min_{\q \in \RR^n} & & \quad \|  M^{\nu}\q \|_{\ell_1} \\ \nonumber
\st  & & \quad  A \cdot \q = \p \\ \nonumber
 & & \quad \langle \1_n,\q \rangle = 1\\ \nonumber
& & \quad \q \geq \0_n. \nonumber  
\end{eqnarray}
Where the entries of the matrix $M^\nu$ are given by $M^{\nu}_{a_xc_z,a^\prime_xb^\prime_yc^\prime_z}= \delta_{a_x,a^\prime_x}\delta_{c_z,c^\prime_z} - f_{a_0,a_1}(\nu)\delta_{c_z,c^\prime_z}$, with $f_{a_0,a_1}(\nu)\equiv q_{a_0,a_1}$, and $\p$ and $\q$ are related by
\begin{eqnarray}
\label{pq}
p(a,b,c|x,y,z)=\sum_{a_x,b_y,c_z} \delta_{a,a_x}\delta_{b,b_y}\delta_{c,c_z}q_{a_0,a_1,b_0,b_1,c_0,c_1}.
\end{eqnarray}
For more details see~\cite{Chaves2015relax}. From the Eq.~\ref{pq}, we can see that the marginals of $\q$ and $\p$ are related by $p(a|x) = \sum_{a_0,a_1}q_{a_0,a_1}$, and we can write the following equations:
\begin{eqnarray}
\label{pqrel}
& & p(0|0) = q_{0,0} + q_{0,1} \\ \nonumber
& & p(1|0) = q_{1,0} + q_{1,1} \\ \nonumber
& & p(0|1) = q_{0,0} + q_{1,0} \\ \nonumber
& & p(1|1) = q_{0,1} + q_{1,1}. \nonumber
\end{eqnarray}
The marginals of $\p$ are known from the expectation va\-lues $\langle A_x\rangle = \sum_{a} p(a|x)$. Using the Eq.~\ref{pqrel} and defining $f_{0,0}(\nu)\equiv\nu$, we have the equations for $f_{a_0,a_1}(\nu)$ in terms of $\langle A_x \rangle$ and $\nu$ as follow:
\begin{eqnarray}
& & f_{0,0}(\nu) = \nu \\ \nonumber
& & f_{0,1}(\nu) = \frac{\langle A_0 \rangle + 1 }{2} - \nu \\ \nonumber
& & f_{1,0}(\nu) = \frac{\langle A_1 \rangle + 1 }{2} - \nu \\ \nonumber
& & f_{1,1}(\nu) = \frac{\langle A_0 \rangle + \langle A_1\rangle + 2 }{2} + \nu. \nonumber
\end{eqnarray}
Once defined the functions $f_{a_0,a_1}(\nu)$, we can write the Eq.~\ref{NBLLP} as a linear program imposing the appropriated constraints.
\begin{eqnarray}
\label{NBLLP2}
\min_{\t \in \RR^n, \q \in \RR^m, \p \in \RR^n} & & \quad \langle  \1_n,\t\rangle \\ \nonumber
\st  & & \quad -\t\leq M^{\nu} \q \leq \t  \\ \nonumber 
& &  \quad A \cdot \q = \p \\ \nonumber
 & & \quad \langle \1_n,\q \rangle = 1\\ \nonumber
 & & \quad \mathrm{M}_{cor}\p = \v_{cor} \\ \nonumber
  & & \quad \sum_{a,b,c}\p(abc|xyz) = 1 \\ \nonumber
 & & \quad \p \geq \0_n \\ \nonumber
& &\quad \q \geq \0_n.\nonumber  
\end{eqnarray}
Where $\v_{cor}=[\langle A_0B_0C_0\rangle ,...,\langle A_1B_1C_1\rangle, \langle A_0\rangle, \langle A_1\rangle]$, for the case with $10$ features, or $\v_{cor} = [\mathrm{I},\mathrm{J},\langle A_0 \rangle, \langle A_1\rangle]$, for the case that we have $4$ features. For the last case, we need to impose the follow constraints $0.25\sum_{x,z} \langle A_xB_0C_z \rangle = \mathrm{I}$ and $0.25\sum_{x,z} (-1)^{x+z} A_xB_1C_z = \mathrm{J}$. All entries of the vector $\v_{cor}$ are random numbers raffled in the interval $[-1, 1]$ . The solution of this linear program can only be given for a fixed known value of $\nu$. In this way, we need to compute the maximum and minimum values of $\nu$, solving two intermediate linear programs:
\begin{eqnarray}
\nu_{min} = \min_{\q\in\RR^m} & & \quad \langle \c,\q \rangle \\ \nonumber
& & \quad A\q=\p \\ \nonumber
& & \quad \q \geq \0_n \\ \nonumber \\
\nu_{max} = \max_{\q\in\RR^m} & & \quad \langle \c,\q \rangle \\ \nonumber
& & \quad A\q=\p \\ \nonumber
& & \quad \q \geq \0_n, \nonumber
\end{eqnarray}
where $\langle \c,\q \rangle = q_{a_0,a_1}$. After that, we sequentially compute the Eq.~\ref{NBLLP2}, for all values of $\nu$ in the range $\nu_{min} \leq \nu \leq \nu_{max}$. In our problem, we have considered $1000$ points from $\nu_{min}$ up to $\nu_{max}$ in order to compute the $M_{BL}^{\nu}$. Thus, minimizing $M_{BL}^\nu \equiv \sum_{a_0,a_1,c_0,c_1} \vert q_{a_0,a_1,c_0,c_1}-f_{a_0,a_1}(\nu)q_{c_0,c_1} \vert$, for a fixed value of $\nu$, is indeed a linear program. In order to verify the non-bi-locality of a given distribution, we need to check if the minimum of $M_{BL}^\nu$ is non-zero for all values of $\nu$ in the allowed range. On the other hand, if we find a value of $\nu$ such that $M_{BL}^\nu = 0$ this is sufficient to show that the distribution is bilocal. 

\end{document}